\documentstyle[12pt,epsf]{article}
\newcommand{\qab} {\mbox{$Q_{ab}$}}
\newcommand{\pq}  {\mbox{$P(q)$}}
\newcommand{\pR}  {\mbox{$P_{R}(q)$}}
\newcommand{\PPP} {\mbox{$P^{(3)}$}}
\renewcommand{\d} {{\rm d}}
\newcommand{\lan} {\langle}
\newcommand{\ran} {\rangle}

\begin{document}

\title{On the origin of ultrametricity}

\author{
Giorgio Parisi$^{(a)}$ and Federico Ricci-Tersenghi$^{(b)}$\\
$^{(a)}$ Dipartimento di Fisica, Universit\`a {\it La  Sapienza}\\
and INFN Sezione di Roma I\\
Piazzale Aldo Moro 2, 00185 Roma (Italy)\\
$^{(b)}$ Abdus Salam ICTP, Condensed Matter Group\\
Strada Costiera 11, P.O. Box 586, 34100 Trieste (Italy)
}

\date{\today}
\maketitle

\begin{abstract}
In this paper we show that in systems where the probability
distribution of the the overlap is non trivial in the infinity volume
limit, the property of ultrametricity can be proved in general
starting from two very simple and natural assumptions: each replica is
equivalent to the others ({\em replica equivalence} or {\em stochastic
stability}) and all the mutual information about a pair of equilibrium
configurations is encoded in their mutual distance or overlap ({\em
separability} or {\em overlap equivalence}).
\end{abstract}

\section{Introduction}

Since its introduction in 1975~\cite{SK} the Sherrington-Kirkpatrick
(SK) model for spin glasses has been one of the major challenges for
the physicists interested in complex systems.

Although it is a mean-field model the exact solution is still not
completely certain.  Nonetheless it is known~\cite{MEPAVI,FISCH_HERTZ}
that in the low-temperature phase the replica symmetry is
spontaneously broken and this makes the solution highly non trivial.
In other words it can be rigorously proved that it is not possible
that the connected correlation function of the spins at different
points goes to zero when the total number of spins $N$ goes to
infinity; consequently the probability distribution of the overlap $q$
(defined below) cannot be a single delta function as happens in usual
model (e.g.\ ferromagnetic model).  The existence of fluctuating
intensive quantities (like $q$) implies that it is not possible that
only one equilibrium state is present in the thermodynamic limit.  As
usual extensive quantities do fluctuate when more than one equilibrium
state is present and consequently we could say, with some abuse of
language (see~\cite{NOI} for a more precise discussion) that in the SK
model, for large values of $N$ more that one equilibrium state is
present.

The presence of many equilibrium states implies that any analytic
solution of the model should tell us something on the nature of these
states, on their relative relations and on the probability at
equilibrium of finding the system in one of these states.  Of course
this information should be of probabilistic nature given the presence
of the quenched disorder in the system.  Using the replica formalism
an Ansatz was proposed almost twenty years ago~\cite{GIORGIO} which
makes some hypothesis on the nature of the states, the most notable
being ultrametricity.  Roughly speaking ultrametricity implies that
the distance among the different states is such that they can be put
in a taxonomic or genealogical tree such that the distance among two
states is consistent with their position on the tree.

This hierarchical Ansatz seems every day more reliable and, although
the physical origin of ultrametricity was not fully evident, it is
widely believed that it provides the correct solution of the SK model.
The ultrametric solution has passed many numerical tests and it is in
agreement with all the known analytical results~\cite{GUERRA,AC}.  It
is quite possible, and in agreement with the numerical simulations,
that the ultrametric organization of the equilibrium configurations is
also present in finite dimensional spin
glasses~\cite{CACCIUTO,INPARU,HARTMANN,UM}.

It is certainly very interesting to find which are the physical
assumptions at the basis of the hierarchical Ansatz for two reasons:
it would be easier to understand if the assumptions make sense also in
the finite dimensional case and it could be easier to prove them of to
extract their consequences.
    
A considerable progress has been done in recent years when it was
realized that one of the main hypothesis at the basis of the
hierarchical Ansatz was stochastic stability: many compulsory
arguments can be given for the validity of stochastic stability and
the correctness of this hypothesis can be directly tested in
experiments measuring the fluctuations and the response to a
perturbation of the appropriate quantities.  In the replica language
{\em stochastic stability} is equivalent to the usual assumption of
{\em replica equivalence} (i.e.\ each replica is equivalent to the
others).

The other pillar of the hierarchical Ansatz turned out to be
ultrametricity. Indeed it can be shown that if we assume stochastic
stability and ultrametricity, the whole hierarchical Ansatz can be
reconstructed~\cite{INPARU}.

The aim of this note is to show that there is a simpler property which
is equivalent to ultrametricity.  This property can be called {\em
separability,} in the replica language, or {\em overlap equivalence}:
it states that all the mutual information about a pair of equilibrium
configurations is encoded in their mutual distance or overlap.  In
other words according to the principle of overlap equivalence any
possible definition of overlaps should not give information additional
to that of the usual overlap.  It is not clear if there are strong
compulsory reasons for assuming overlap equivalence, however, the
results presented here show that the hierarchical Ansatz is certainly
the simplest one that we may think for a stochastically stable system
with many equilibrium states: any other possible proposal should
include the presence of at least two inequivalent definitions of
distance.

These two assumptions, stochastic stability and overlap equivalence,
are quite general and can be applied to many other systems beyond the
SK model.  A direct test or an analytic proof of the validity of both
properties would have direct implications on the validity of the
ultrametric solution.  Moreover as we have already remarked the
results that we are going to present in this paper is interesting
because they show the root of ultrametricity: ultrametricity is the
unique possibility we have if we stay within the simple framework
where stochastic stability and separability hold.

It should be clear that the whole discussion applies to systems in
which the overlap fluctuates also when the volume is very large and
consequently replica symmetry is broken.  For systems in which the
overlap does not fluctuates and replica symmetry is exact we have
nothing to say (ultrametricity is satisfied, but in a trivial way).
It should also be clear that the arguments presented here cannot be
used to argue if replica symmetry breaking happens or not in a
particular system.  We do not discuss here the criticism that have
been done to the replica approach by~\cite{FH, NS} (who cast some
doubts on the viability of replica symmetry breaking in finite
dimensional systems): the reader may find a quite long discussion
in~\cite{NOI}.

In the second section we will recall the replica formalism, in the
third section we will present our assumptions.  Next in the forth
section we will present our main results on the relation among overlap
equivalence and ultrametricity.  Finally we will present our
conclusions.  Some of the arguments needed to show that overlap
equivalence implies separability are presented in Appendix I and a
part of the tedious algebra we have to do to reach the results is
confined in appendix II.

\section{The replica formalism}

In this paper we make use of the replica formalism (we address the
reader to Refs.~\cite{MEPAVI, FISCH_HERTZ, PAR_JSP, PAR_98} for an
introduction on the issue).  Let for simplicity restrict the
discussion to systems with quenched random disorder in the
Hamiltonian.  A similar discussion can be done for system without
quenched disorder (like structural glasses) if we substitute the
average over the quenched disorder with the average over the size of
the system.

We consider a system with $N$ spins characterized by an Hamiltonian
$H_{J}(\sigma)$ (where $J$ represents the quenched disorder).  We
define $P_{J}(q)$ to be the probability distribution of finding two
equilibrium configurations $\sigma$ and $\tau$ at the same inverse
temperature $\beta$ with overlap $q$, the overlap $q$ being defined as
\begin{equation}
q = \frac1N \sum_{i=1,N} \sigma_{i} \tau_{i} \quad .
\end{equation}

Let us assume that, in the large $N$ limit, the function $P_{J}(q)$
does not becomes a delta function (otherwise we have nothing
interesting to say) and therefore the function $P_{J}(q)$ has a non
trivial shape also for large $N$.  When this happens, we are
interested to find out the probability distribution of the function
$P_{J}(q)$ in the limit where $N$ goes to infinity.  For a given value
of $N$ different choices of $J$ may produce different functions
$P_{J}(q)$ and we can introduce the functional ${\cal P}_{N}(P)$ as
the probability distribution of $P_{J}(q)$ at fixed $N$ (we assume
that the $J$ have a given probability distribution). Eventually we
would like to know
\begin{equation}
{\cal P}_{\infty}(P) \equiv \lim_{N \to \infty}{\cal P}_{N}(P) \quad .
\end{equation}
We are also interested to control the behavior of the probability
distribution of the mutual overlaps among three or more equilibrium
configurations (e.g.\ the probability
$P^{12,23,31}_{J}(q_{12},q_{23},q_{31})$ which will be properly
defined later).

The origin of our interest in the probability distribution of the
overlap is due to the fact that they control many others physical
properties of the systems: e.g.\ in some models one
finds~\cite{PARIRU} that the magnetic susceptibility is given by
$\chi=\beta(1-\overline{\lan q \ran})$ where $\overline{\lan q \ran}$
is the average over $J$ of the equilibrium expectation value of $q$.

In the replica formalism the behavior of the these probability
distributions is encoded by an $n \times n$ symmetric matrix $Q$ in
the limit of $n \to 0$ (taken after the analytical continuation of $n$
from integer to real values).  The limiting matrix depends on all the
matrices with any value of $n$ and so the general solution has an
infinite number of parameters and the analytical continuation of the
matrix $Q$ is, in general, dependent on an extremely high number of
parameters.  This is quite natural as far as the matrix $Q$ encodes
the properties of the functional which controls the probability
distribution of finding for a random $J$ a set of probability
distributions for the overlaps (i.e.\ $P_{J}(q)$,
$P^{12,23,31}_{J}(q_{12},q_{23},q_{31})$, \ldots).
 
In the hierarchical Ansatz the $n$ replicas are divided into many
groups of equal sizes, such that, if the replica indices $a$ and $b$
belong to the same group, then \qab\ has a higher value than whether
$a$ and $b$ are in different groups.  The groups are then divided in
subgroups and so on for an infinite number of times.  This kind of
solution can be summarized in an infinite set of parameters (the size
of the group and the value of the overlap at each level).  In the
limit $n \to 0$ these parameters can be conveniently represented by a
function \pR\ defined for $q \in [0,1]$, where \pR\ is the probability
of finding in the matrix $Q$ an element of value $q$.  To every
ultrametric matrix $Q$ corresponds one and only one probability
distribution function \pR.

In the paramagnetic phase all the elements of $Q$ are equal and the
function \pR\ is a delta function.  In the spin glass phase the
elements \qab\ take different values and the \pR\ acquires a finite
width.

The relation of this function with the probability distribution
function of the overlap is
\begin{equation}
\pR = P(q) \equiv \lim_{N\to\infty} \overline{P_{J}(q)} \quad ,
\label{eq:a}
\end{equation}
where the bar denotes the average over $J$ at fixed $N$. The equality
of the two functions \pR\ and $P(q)$ is one of the many relations
among probability distribution functions of the overlaps and the
matrix $Q$.

More complicated probability distribution functions (pdf) can be
defined, considering the joint probability of more than one overlap.
For example a crucial role is played by the joint pdf of 3 real
replicas $P^{12,23,31}(q_{12},q_{23},q_{31})$, defined as the
$J$-average of $P_{J}^{12,23,31}(q_{12},q_{23},q_{31})$ where
$\sigma^{1}$, $\sigma^{2}$ and $\sigma^{3}$ are three equilibrium
configurations and
\begin{equation}
q_{\alpha,\beta} = \frac1N \sum_{i=1,N} \sigma^{\alpha}_{i}
\sigma^{\beta}_{i} \quad ,
\label{eq:defq}
\end{equation}
with the indices $\alpha$ and $\beta$ running from 1 to 3.

If ultrametricity holds, this probability distribution has the
following property
\begin{equation}
P^{12,23,31}(q_{12},q_{23},q_{31}) = 0 \quad ,
\label{eq:UL}
\end{equation}
as soon as the ultrametricity relations 
\begin{eqnarray}
q_{12} \ge \min(q_{23},q_{31}) \quad , \nonumber \\
q_{23} \ge \min(q_{31},q_{12}) \quad , \label{eq:DIS} \\
q_{31} \ge \min(q_{12},q_{23}) \quad , \nonumber \,
\end{eqnarray}
are not satisfied.  The UM property can be easier understood by a
geometrical picture.  Given three configurations, that is three points
in the configurational space, UM implies that they can be the vertices
of two kind of triangles only: equilateral or isosceles, with the two
equal edges larger than the third one.  The other kind of isosceles
triangles, together with the scalene triangles, can not be obtained
with any tern of equilibrium configurations, if these configurations
are organized in an UM fashion.

This property, which is satisfied in the hierarchical Ansatz, has
rather strong consequences.

Firstly, as far as probabilities cannot be negative, the previous
relations implies that also for any $J$ (with probability one) we have
that $P_{J}^{12,23,31}(q_{12},q_{23}, q_{31})=0$ as soon as the
relations in Eq.(\ref{eq:DIS}) are not satisfied.  This result has the
consequence that any equilibrium configuration can be assigned (for
fixed $J$) to a leave of a tree constructed in such a way that the
overlap among two equilibrium configuration is related to the distance
of the two configurations on the tree.  Ultrametricity implies for
example that, if two equilibrium configurations 1 and 2 are at overlap
$q_{12}>q$, any equilibrium configuration 3 such that $q_{13}>q$
satisfies also the relation $q_{23}>q$.  Ultrametricity is very
interesting because it implies that many pdf of more than three
overlaps are zero in a wide region and reduces the whole problem to
the construction of the statistical properties of the aforementioned
tree.

Moreover, if stochastic stability is valid, the ultrametricity
completely determines the $P^{12,23,31}$ given the \pq. One can show
that~\cite{INPARU}
\begin{eqnarray}
P^{12,23,31}(q,q',q'') = A(q)\,\delta(q-q')\,\delta(q-q'') +
B(q,q')\,\theta(q-q')\,\delta(q'-q'') + \nonumber \\
B(q',q'')\,\theta(q'-q'')\,\delta(q''-q) +
B(q'',q)\,\theta(q''-q)\,\delta(q-q') \quad ,
\end{eqnarray}
where
\begin{equation}
A(q) = \frac12 P(q) \int_0^q \d q' P(q') \quad ,
\end{equation}
and
\begin{equation}
B(q,q') = \frac12 P(q) P(q') \quad .
\end{equation}
In other words stochastic stability and ultrametricity allow us to
obtain all the pdf of the overlap starting from the knowledge of the
function \pq.

\section{The assumptions}

It is clear that it is extremely difficult to arrive to some general
conclusions on these probability distributions without doing extra
assumptions.  We now show that two rather simple assumptions: replica
equivalence (or equivalently stochastic stability) and separability
are giving very strong constraints.

Let us firstly consider replica equivalence as formulated in replica
formalism.

As we have already stated the properties of the probability
distribution of the overlaps can be obtained in terms of the matrix
\qab.  Even in the low-temperature phase, when the matrix elements
\qab\ are not constant, we may expect no physical difference between
the replicas (which have been introduced as a mathematical trick).
Replica equivalence states that the observable which involve only one
replica are replica symmetric, i.e.\ the assume the same value.  For
example replica equivalence implies that we must have that
\begin{equation}
\sum_b f(Q_{ab})
\label{re}
\end{equation}
does not depend on $a$.

Replica equivalence is equivalent to the stochastic stability property
introduced by Guerra~\cite{GUERRA} and Aizenman and Contucci~\cite{AC}
which is valid under general conditions, i.e.\ if we introduce an
arbitrary small random long range Hamiltonian (see~\cite{GUERRA} for a
more careful discussion).

Eq.(\ref{re}) implies that each line (column) of the matrix $Q$ is a
permutation of the other lines (columns).  Moreover it has interesting
consequences: with some algebra the following equalities can be proven
\begin{eqnarray}
P^{12,13}(q_{12},q_{13}) = \frac12 P(q_{12}) P(q_{13}) +
\frac12 P(q_{12}) \delta(q_{12}-q_{13}) \quad , \nonumber \\
P^{12,34}(q_{12},q_{34}) = \frac23 P(q_{12}) P(q_{34}) +
\frac13 P(q_{12}) \delta(q_{12}-q_{34}) \quad . \label{GUE_REL}
\end{eqnarray}

The proof of these equations can be done by recalling some relations
among the matrix \qab\ and the probability functions
\begin{eqnarray}
P^{12,13}(q_{12},q_{13}) = \lim_{n\to 0} {\sum'_{a,b,c=1,n}
Q_{a,b}Q_{a,c} \over n(n-1)(n-2)} \ , \nonumber \\
P^{12,34}(q_{12},q_{34}) = \lim_{n\to 0} {\sum'_{a,b,c,d=1,n}
Q_{a,b}Q_{d,c} \over n(n-1)(n-2)(n-3)}\ , \label{PRIMA}
\end{eqnarray}
where the primed sum is done over all different replica indices.  Let
us denote
\begin{equation}
q^{(k)} = -\sum_{b=1,n}q_{a,b}^{k}
\end{equation}
The sum does not depend on $a$ as consequence of replica equivalence.
It is also evident that
\begin{equation}
\sum_{a,b=1,n}q_{a,c}^{k_{1}} q_{b,d}^{k_{2}}=q^{(k_{1})}q^{(k_{2})}
\end{equation}

If we now look to the consequences of the previous equation and we use
the relations in Eq.(\ref{PRIMA}) both for $a=b$ and $a\ne b$, we
obtain the two relations in Eq.(\ref{GUE_REL}).

Identical relations have been proven by Guerra~\cite{GUERRA}, using
stochastic stability.  We can very safely assume that they must be
valid in any scheme of replica symmetry breaking.  Eq.(\ref{GUE_REL})
determine all the joint pdf of 2 real replicas in terms of the \pq.
The consequences of stochastic stability have been lengthy discussed
in~\cite{PAR_98,CORTO,LUNGO}. In the nutshell stochastic stability
implies that the system is a generic random system and it does not
have any special properties: its properties are smooth functions of
any external random perturbation.

The second assumption we made is the separability (also know as
non-degeneracy) of the matrix $Q$~\cite{PAR_98}, which correspond to
the following statement.  Let us consider all the matrices which can
be generated from the matrix $Q$ in a permutational covariant
fashion. Some example are
\begin{equation}
Q_{ab}^k \quad , \quad \sum_c Q_{ac}Q_{cb} \quad , \quad \sum_{c,d}
Q_{ac}Q_{ad}Q_{cd}Q_{cb}Q_{db} \quad . \label{SET}
\end{equation}
Separability states that, if we take two pair of indices ($ab$ and
$cd$), we have that
\begin{equation}
Q_{ab} = Q_{cd} \Longrightarrow M_{ab} = M_{cd}
\label{SEPA}
\end{equation}
where $M$ is a generic matrix of the set generated by the rules shown
in Eq.(\ref{SET}).  In other words pairs of indices which have
different properties have different values of the overlap.  It means
that we can classify a pair of replicas in terms of their mutual
overlap~\cite{LUNGO} and that no finer classification of their mutual
properties is possible.

The physical meaning of separability can be understood if we introduce
another concept, the overlap equivalence. Let us consider an arbitrary
local observable $O_i(\sigma)$. Simple examples of such an observable
are
\begin{eqnarray}
O_{i} = \sum_k A_{i-k} \sigma_k ,\nonumber \\
O_{i} = \sum_{k,l} B_{i-k,i-l} \sigma_k \sigma_l ,\\
O_{i}=\sum_{k}\sigma_i\sigma_k J_{i,k} \nonumber
\end{eqnarray}
where $A$ and $B$ are appropriate functions (e.g.\ they decrease
sufficiently fast at infinity).  Many more complex choices of the
local observable $O$ can be constructed, for example those involving
more than two spins.

For any choice of the operator $O$ we could define a generalized
overlap~\cite{FACING}:
\begin{equation}
q_O = \frac1N \sum_{i=1,N} O_i(\sigma) O_i(\tau) \quad .
\end{equation}

In the hierarchical Ansatz it turns out~\cite{BACHAS} that for any
reasonable choice of the observable $O$, $q_{O}$ is a function of $q$.
In other words when we change the two equilibrium configurations and
the couplings $J$ the values of $q$ and $q_{O}$ fluctuates also for
very large $N$, while the value of $q_{O}$ restricted on those pairs
of configurations with a fixed value of $q$ does not fluctuate when
$N$ goes to infinity (that is a scattered plot of $q$ and $q_{O}$
should collapse on a curve in the limit of large $N$).  In other words
overlap equivalence implies that in the case where replica symmetry is
broken and all overlaps fluctuate in the usual thermodynamical
ensemble, these fluctuations disappear in the fixed $q$ ensemble.

In other words overlap equivalence states that for a system composed
by two replicas the overlap is a good, complete order parameter in the
same way as the magnetization is for ferromagnetic systems.  If we
stay in the usual thermodynamic ensemble, there are many quantities
that fluctuates also at large distances, however if we consider the
restricted ensemble where the order parameter takes a given value, all
fluctuations at large distance disappear and the connected correlation
functions go to zero at infinity.  This happens only if the order
parameter has been chosen in such a way to carry enough information:
in a ferromagnetic Heisenberg model the order parameter must be the
three component vector of the magnetization, one or two components of
the magnetization would be not enough to fully characterize the state
of the system in case of a spontaneous magnetization.

A direct check of overlap equivalence can be done in usual numerical
simulations and it would be very interesting to see the results.

This property is called overlap equivalence because it states that all
possible definitions of the overlap are equivalent and there is an
unique correspondence among the values of the different overlaps.

It is clear that the overlap equivalence is a very strong
simplification.  In general we could have that the mutual relations
among two equilibrium configurations are characterized by a large,
possibly infinite set of independent overlaps and therefore their
mutual relations are characterized by a large (or infinite) set of
parameters.  The property of overlap equivalence implies a much
simpler situation, where only one parameter (the overlap $q$)
characterizes the mutual relations among two equilibrium
configurations.

We can argue that separability is the way to code overlap equivalence
in the replica formalism.  Both properties state that once the overlap
among two {\sl objects} is fixed, all the mutual relations among the
two objects are also fixed.  The difference among these two statements
is that in the case of replica equivalence the two objects are
equilibrium configurations while in the case of separability the two
objects are replicas.  The identification of separability with overlap
equivalence is quite natural because the structure of the matrix $Q$
in replica space mirrors the structure of the mutual overlaps of
equilibrium configurations.  In appendix I we present some more
detailed considerations which points toward the correctness of the
identification of these two properties, however a more general and
formal proof of this statement would be welcome.

It is interesting to note that in the simplest model leading to
ultrametricity, i.e.\ a branching random process in the infinite
dimensional space the condition of overlap equivalence is
satisfied~\cite{RATOVI}.  Indeed if we consider a random vector
$x_{\alpha}$ in a finite dimensional space (of dimension $N$) the
quantities $x_{\alpha}^{2}$ convey different information when $\alpha$
changes from 1 to $N$ and can be used as different measures of the
distance.  On the other end, when $N$ goes to infinity at fixed
$x^{2}\equiv\sum_{\alpha=1,N}x_{\alpha}^{2}$, thanks to the rotational
invariance, we have that for each $\alpha$, $\lan x_{\alpha}^{2} \ran
= {x^2 \over N} \to 0 $.  Then if we introduce generalized distances
parameterized by $\lambda$ (where $0<\lambda\le 1$) and defined as
\begin{equation}
x^{2}_{\lambda} \equiv \lambda^{-1} \sum_{\alpha=1,\lambda N}
x_{\alpha}^{2} \quad ,
\end{equation}
it is easy to check that with probability 1 (if the probability
distribution is rotational invariant) in the limit $N \to \infty$
\begin{equation}
x^2_\lambda = x^2_1 = x^2 \quad .
\end{equation}
Therefore in this simply model overlap equivalence is automatically
satisfied.

Let us consider what happens in the usual hierarchical Ansatz.  In
this case, when replica symmetry is broken, there is a subgroup of the
group of permutations that commutes with the matrix $Q$.  Let us
consider the orbits in the space of pairs of indices.  It can be
checked that the values of the elements of the matrix $Q$ and of any
matrix derived using the rules in Eq.(\ref{SET}) are constant of the
orbits and that different values of $q$ do correspond in general to
different orbits.  Moreover it can be checked that both separability
and overlap equivalence hold in this case.

Maybe the simplest non-trivial example of a non-ultrametric system is
given by the union of separately ultrametric system with a non-trivial
distribution of the overlaps~\cite{FRPAVI}.  It is easy to check that
$q=\lambda q_{1}+(1-\lambda) q_{2}$ does not satisfy the ultrametric
condition also if $q_{1}$ and $q_{2}$ do satisfy it.  However it is
clear that in this example $q_{1}$ and $q_{2}$ are generalized
overlaps which are not function of $q$.  Both ultrametricity and
overlaps equivalence disappear at the same time~\footnote{In the last
example both ultrametricity and stochastic stability are violated.
There are no known examples of stochastically stable states which are
not ultrametric.  As far as we know, it is still possible that
stochastic stability implies ultrametricity.}.

The separability condition is extremely powerful in determining the
expectation values of higher order moments of the probability
distribution.  Let us study a simple example and let us consider two
matrices $M$ and $R$ constructed with the rules in Eq. (\ref{SET}). It
is evident that we can write
\begin{eqnarray}
\sum_b Q_{ab}^k M_{ab} = \sum_{b}\int \d q \delta(q-Q_{ab}) Q_{ab}^k
M_{ab}= \int \d q P(q) M(q) q ^k \quad , \nonumber\\
\sum_b Q_{ab}^k R_{ab} = \sum_{b}\int \d q \delta(q-Q_{ab}) Q_{ab}^k
R_{ab} \int \d q P(q) R(q) q ^k \quad , \label{DUE}
\end{eqnarray}
where we have defined 
\begin{equation}
P(q)=\sum_{b}\delta(q-Q_{ab}) \quad .
\end{equation}
Indeed separability implies that the matrix elements $M_{ab}$ and
$R_{ab}$ are constant in the region where $Q_{ab}=q$ and their value
is denoted $M(q)$ and $R(q)$ respectively. In the same way we have
that:
\begin{equation}
\sum_b Q_{ab}^k M_{ab} R_{ab} = \int \d q P(q) M(q) R(q) q^k \quad .
\label{PROD}
\end{equation}
Therefore separability implies that quantities like those in
Eq.(\ref{PROD}) can be computed from the knowledge of those in
Eq.(\ref{DUE}).

If we introduce the functions $P_{M}(q)$, $P_{R}(q)$ and $P_{MR}(q)$
such that
\begin{eqnarray}
\sum_b Q_{ab}^k M_{ab} = \int \d q P_{M}(q) q ^k \quad , \nonumber \\
\sum_b Q_{ab}^k R_{ab} = \int \d q P_{R}(q) q ^k \quad , \\
\sum_b Q_{ab}^k M_{ab} R_{ab} = \int \d q P_{MR}(q) q ^k \quad , \nonumber
\end{eqnarray}
the previous equations implies that
\begin{equation}
P_{M}(q)=P(q) M(q), \ \ P_{R}(q)=P(q) R(q), \ \ P_{MR}(q)=P(q)
M(q)R(q).
\end{equation}
The last equation can also be written as
\begin{equation}
P_{MR}(q)={P_{M}(q)P_{R}(q) \over P(q)}\quad .
\end{equation}
If we apply the previous formula to the case where $M$ and $R$ have
the form
\begin{eqnarray}
M_{ab} = \sum_c Q_{ac}^{k_1} Q_{cb}^{k_2} \quad ,\\
R_{ab} = \sum_c Q_{ac}^{k_3} Q_{cb}^{k_4} \quad ,\,
\end{eqnarray}
and we consider all the possible values of the $k$'s, we find (after
separating the contributions where some of the indices are equal) the
rather surprising formula
\begin{eqnarray}
3\,P^{12,13,32,24,41}(q,q_1,q_2,q_3,q_4) = \delta(q_1-q_4)
\delta(q_2-q_3) P^{12,23,31}(q,q_1,q_2) + \nonumber \\
+2 {P^{12,23,31}(q,q_1,q_2) P^{12,23,31}(q,q_3,q_4) \over P(q)} \quad .
\label{WOH}
\end{eqnarray}
Similar results can be obtained for other probability distributions
with more overlap.

Eq.(\ref{WOH}) is particular interesting because integrating over $q$
it implies that
\begin{eqnarray}
3\,P^{13,32,24,41}(q_1,q_2,q_3,q_4) =
\frac12 \delta(q_1-q_4) \delta(q_2-q_3) \left[P(q_1) P(q_2) +
\delta(q_{1}-q_{2}) P(q_{2})\right] + \nonumber \\
+2 \int \d q {P^{12,23,31}(q,q_1,q_2) P^{12,23,31}(q,q_3,q_4) \over
P(q)} \quad .
\label{MAIN}
\end{eqnarray}

The previous equation is remarkable not only because it give the full
expression of the probability with four overlap in terms of the
probability with three overlaps but also because it puts hard
constraints on the possible values of the function
$P^{12,23,31}(q,q_1,q_2)$.  Indeed the l.h.s.\ of Eq.(\ref{MAIN}) is
by definition invariant under cyclic permutation of the $q$'s, while
the r.h.s.\ of the same equation is not invariant for a generic choice
of the function $P^{12,23,31}$.

What is the form of the generic function $P^{12,23,31}$ that satisfy
Eq.(\ref{MAIN})?  We will argue in the next section that it must be
ultrametric.

\section{Results}

Our problem is now that of finding the most general matrix $Q$ (or
equivalently the most general probability distribution) compatible
with the replica equivalence [and then with Guerra's relations,
Eq.(\ref{GUE_REL})] and with separability [and then in particular with
Eq.(\ref{MAIN})].  We will show that the most general matrix is the
ultrametric one.

We consider the case when few values ($k=3$, $k=4$ or $k=5$) are
allowed for the matrix elements.  The generalization to more than 5
values is straightforward and we hope that our conclusions will still
be valid for a generic \pq\ which have a continuous distribution of
possible values.

When the overlap (or the matrix elements \qab) can take only $k$
different values the function \pq\ is the sum of $k$ delta functions
\begin{equation}
\pq = \sum_{i=1}^k p_i\ \delta(q-q_i) \quad ,
\end{equation}
where the weights $p_i$ are, by definition, positive and such that
$\sum_i p_i = 1$, and the values $q_i$ are different.  Also the joint
pdf of 3 overlaps $P^{12,23,31}$ (that hereafter we will call \PPP\
for brevity) is the the sum of $k^3$ delta functions on the points
$(q_i,q_j,q_l)$ with $i,j,l=1,\ldots,k$ and so we should give the
$k^3$ weights $p_{ijl}$ in order to determine \PPP.

We can lower the number of free parameters $p_{ijl}$ using some
symmetries and the Guerra relations.  The weight of the tern
$(q_i,q_j,q_l)$ must be the same of any permutation of it, i.e.\
$p_{122}=p_{212}=p_{221}$.  Then the number of really independent
parameters in \PPP\ is $k(k+1)(k+2)/6$.  More relations between the
$p_{ijl}$ can be obtained exploiting the following equation, which is
based essentially on the first Guerra's relation
\begin{equation}
\int \d q \PPP (q,q_1,q_2) = P^{12,23}(q_1,q_2) = \frac12 P(q_1)
P(q_2) + \frac12 P(q_1) \delta(q_1-q_2) \quad .
\end{equation}
These are $k(k+1)/2$ relations that lower the degrees of freedom of
\PPP\ to $(k-1)k(k+1)/6$.

Then we have to determine the values of these $(k-1)k(k+1)/6$
parameters which are compatible with Eq.(\ref{MAIN}).

\subsection{Three overlaps ($k=3$)}

To fix the ideas, let us write down some formul\ae\ for the easier
case ($k=3$) where we have 27 parameters $p_{ijl}$: $p_{111}, p_{112},
p_{113}, p_{121}, \ldots, p_{333}$.  The symmetries imply that
\begin{equation}
\left\{
\begin{array}{l}
p_{112} = p_{121} = p_{211} \quad ,\\
\ldots \\
p_{123} = p_{132} = p_{213} = p_{231} = p_{312} = p_{321} \quad ,
\end{array}
\right.
\end{equation}
while the Guerra's relations imply some equalities like
\begin{equation}
\left\{
\begin{array}{l}
\sum_j p_{11j} = \frac12 p_1^2 + \frac12 p_1 \quad ,\\
\ldots \\
\sum_j p_{12j} = \frac12 p_1 p_2 \quad ,\\
\ldots
\end{array}
\right.
\end{equation}
We end with only 4 free parameters ($s,a_{32},a_{31},a_{21}$):
\begin{equation}
\begin{array}{l}
p_{321} = s \\
p_{332} = a_{32} \\
p_{331} = a_{31} \\
p_{221} = a_{21} \\
p_{322} = p_3 p_2/2 -a_{32} -s \\
p_{311} = p_3 p_1/2 -a_{31} -s \\
p_{211} = p_2 p_1/2 -a_{21} -s \\
p_{333} = p_3(1+p_3)/2 -a_{31} -a_{32} \\
p_{222} = p_2(1+p_2)/2 -p_3 p_2/2 +a_{32} -a_{21} +s \\
p_{111} = p_1(1+p_1)/2 -p_3 p_1/2 -p_2 p_1/2 +a_{31}
+a_{21} +2 s
\end{array}
\label{GRANDE}
\end{equation}
The way we have ordered the probabilities is meaningful: we call $s$
the weight of the scalene triangle (which is forbidden in the
ultrametric solution) and we call $a_{32},a_{31},a_{21}$ the weights
of the isosceles triangles (which are also forbidden in the UM Ansatz
if we assume the overlap ordering $q_1<q_2<q_3$).

If we do not fix any order in the values of the $q_i$, we have to keep
in mind that, whether we exchange two of the overlap values, the
forbidden isosceles triangle changes.  For example, if $q_1<q_2<q_3$
then ultrametricity implies $p_{332} = p_{331} = p_{221} = 0$, while
when we reverse the second inequality, i.e.\ $q_1<q_3<q_2$, we have
that $p_{322} = p_{331} = p_{221} = 0$.  Then we note that
\begin{equation}
s=a_{32}=a_{31}=a_{21}=0 \quad \Longrightarrow \quad \mbox{UM} \quad ,
\end{equation}
while the reversed implication is not true, because UM also holds for
different parameters values, e.g.\ $s=a_{31}=a_{21}=0$ and $a_{32}=p_3
p_2/2$ (that corresponds to the ordering $q_1<q_3<q_2$).

For a generic $k$ we have $k(k-1)(k-2)/6$ scalene parameters $s_i$
which must be all identically zero in order that UM hold ($\{s_i=0\}
\Longleftrightarrow \mbox{UM}$), while the $k(k-1)/2$ isosceles
parameters $a_{ij}$ must be zero or $a_{ij}=\frac12 p_i p_j$,
depending on the order of $q_i$ and $q_j$.

We will now use Eq.(\ref{MAIN}) to determine the values of all these
parameters.  The l.h.s.\ of Eq.(\ref{MAIN}) is invariant under cyclic
permutations of the four overlaps.  This allows us to obtain useful
relations simply taking two of these equations (the second one with
the overlaps cycled with respect to the first one) and equating the
right hand sides.  The number of non-trivial equations we can obtain
in this way is large enough to fix all the parameters.

In the particular case of $k=3$ we have that all the non-trivial
equations are equal (this is highly fortuitous) and read
\begin{eqnarray}
-2 a_{32} a_{31} p_2 p_1 +2 a_{32} a_{21} p_3 p_1 -2 a_{31} a_{21}
p_3 p_2 +a_{31} p_3 p_2^2 p_1 +2 a_{32} p_3 p_1 s + \nonumber\\
-2 a_{32} p_2 p_1 s -2 a_{31} p_2 p_1 s +p_3 p_2 p_1 s
-p_3^2 p_2 p_1 s +2 p_3 p_2 s^2 +2 p_3 p_1 s^2 = 0
\label{BRUTTA}
\end{eqnarray}
Using the relations that comes from the sixth equation in
(\ref{GRANDE})
\begin{equation}
a_{31} = \frac12 p_3 p_1 -s -p_{311} \quad ,
\end{equation}
we can write Eq.(\ref{BRUTTA}) as
\begin{equation}
E_0 +2 (a_{32} p_3 p_1 +a_{21} p_3 p_2 +p_{311} p_2 p_1) s +2 (p_3 p_2
+p_3 p_1 +p_2 p_1) s^2 = 0 \quad ,
\label{BELLA}
\end{equation}
where in $E_0$ we put all the terms that survive once we set $s=0$.
The coefficients of $s$ and $s^2$ are positive defined (thanks to the
positiveness of all the probabilities) and $E_0$ in non-negative (as
we will show in a while).  Then Eq.(\ref{BELLA}) is equivalent to
\begin{equation}
\left\{
\begin{array}{l}
s = 0 \quad ,\\
E_0 = a_{31} p_3 p_2^2 p_1 -2 a_{32} a_{31} p_2 p_1 -2 a_{31} a_{21}
p_3 p_2 +2 a_{32} a_{21} p_3 p_1 = 0 \quad .
\end{array}
\right.
\label{S0}
\end{equation}
As a first result we obtain that scalene triangles are completely
forbidden.

\begin{figure}
\begin{center}
\leavevmode
\epsfxsize = 0.4 \textwidth
\epsffile{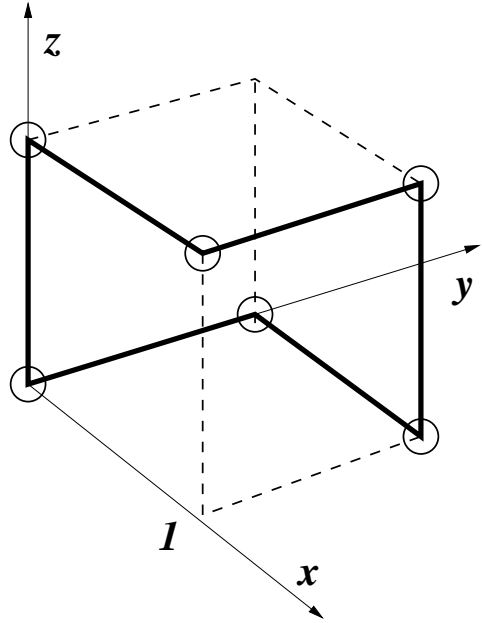}
\end{center}
\caption{Schematic representation of the properties of the symbol
$((x;y,z))$, it takes non-negatives value in all the unitary cube and
it is zero only on the bold edges.}
\label{fig}
\end{figure}

Let us now introduce the following symbol
\begin{equation}
((x;y,z)) \equiv x -x y -x z +y z = x(1-y)(1-z) +(1-x)y z \quad .
\end{equation}
For $x,y,z\in[0,1]$ we have that $((x;y,z))\ge0$ and the equality
$((x;y,z))=0$ only holds on 6 of the 12 edges of the cube (those in
bold face in Fig.~\ref{fig}).

If we introduce the new parameters
\begin{equation}
a'_{ij} \equiv \frac{2 a_{ij}}{p_i p_j} \quad ,
\end{equation}
that belong to the range $[0,1]$ thanks to the positiveness of the
probabilities, then the second equality in Eq.(\ref{S0}) can be
rewritten in a very compact form as
\begin{equation}
((a'_{31};a'_{32},a'_{21})) = 0 \quad .
\label{CORTA}
\end{equation}
This form makes clear that $E_0$ is non-negative, as we claimed above.

Eq.(\ref{CORTA}) is not so stringent as ultrametricity would, but the
deviations from UM are small.  In fact, in the cube
$a'_{32},a'_{31},a'_{21}\in[0,1]$ (see Fig.~\ref{fig}) strict UM only
holds on the vertices marked by a circle.  While Eq.(\ref{CORTA}) is
satisfied along the bold lines too.

For example, on the segment $a'_{32}=a'_{31}=0$ and $0 < a'_{21} < 1$
they seem to co-exist non-zero probabilities $p_{221}$ and $p_{211}$
and it would be a small violation of UM.  However we know that
$a'_{32}=a'_{31}=a'_{21}=0$ correspond to the ordering $q_1<q_2<q_3$,
while $a'_{32}=a'_{31}=0$ and $a'_{21}=1$ correspond to $q_2<q_1<q_3$.
Then we believe that the points on the segment between this two UM
points corresponds to the case $q_1=q_2<q_3$, when there is no
difference between $p_{221}$ and $p_{211}$ (but we can not still prove
it).

In conclusion, in the case with $k=3$ overlaps, we have that the
scalene triangle and two of the three ``wrong'' (in the UM sense)
isosceles triangles are forbidden.  As we will see below, the UM
violations become smaller and smaller as $k$ is increased.

\subsection{Four or more overlaps ($k \ge 4$)}

In this section we would like to sketch how the information we need
about \PPP\ can be systematically derived from Eq.(\ref{MAIN}).  To
make this section more readable, the formul\ae\ relatives to the cases
$k=4,5$ will be presented in Appendix II.  The method we use to obtain
the results does not depend on $k$ and so we will be able to
generalize our findings to whichever \pq\ that is the sum of a finite
number of delta functions.

The many equations derivable from Eq.(\ref{MAIN}) can be divided into
three classes: those with 2, 3 and 4 different values of the overlaps.
These equation are not independent: those with 2 (respectively 4)
different overlaps can be expressed as the sum (resp. difference) of
those with 3 overlaps.

There are many ways of solving the equations.  Here we present the
simplest one we were able to find: we have to consider only the
equations with 2 and 3 overlaps, i.e.\ those which respectively
correspond to the equalities
\begin{equation}
P^{12,23,34,41}(q_i,q_j,q_j,q_i) - P^{12,23,34,41}(q_i,q_i,q_j,q_j) =
0 \quad , \label{DUEQ}
\end{equation}
and
\begin{equation}
P^{12,23,34,41}(q_i,q_i,q_j,q_l) - P^{12,23,34,41}(q_i,q_j,q_l,q_i) =
0 \quad . \label{TREQ}
\end{equation}
Each one of these equations can be identified giving a pair or a tern
of numbers: $(i,j)$ or $(i,j,l)$.  The l.h.s.\ of these equations will
be called respectively $E^{(i,j)}$ and $E^{(i,j,l)}$, for brevity.

Our demonstration follows two steps: first we show that the equations
of the kind of Eq.(\ref{DUEQ}) can be solved only if all the scalene
parameters $p_{ijl}$ (with $i,j,l$ different) are zero, then we find
all the solutions for the simplified set of equations corresponding to
Eq.(\ref{TREQ}).  Our demonstration is essentially based on the
non-negativity of the $E^{(i,j)}$ expressions and on the properties of
the double-parenthesis symbol, previously introduced.

First of all we note (see Appendix II) that when we set to zero all
the scalene parameters $p_{ijl}$ (with $i,j,l$ different), every
$E^{(i,j)}$ becomes the sum of some double-parenthesis symbols, and so
it is non-negative.  Moreover in some of the $E^{(i,j)}$ expressions
all the scalene parameters have positive defined coefficients, and
then we should set them to zero in order to solve the equation,
$E^{(i,j)}=0$.  In Appendix II we present a possible way of choosing
the $E^{(i,j)}$ expressions in order to systematically set to zero all
the scalene parameters.

Once the scalene parameters have been set to zero, we prefer working
with Eq.(\ref{TREQ}), because each equation identified by $(i,j,l)$
takes a very simple form:
\begin{equation}
((a'_{il};a'_{ij},a'_{jl})) = 0 \quad ,
\label{FINAL}
\end{equation}
where we choose the indexes such that $q_i\!<\!q_j\!<\!q_l$.  In
general, given 3 different overlap values, we can easily write down
the corresponding equation [of the kind of Eq.(\ref{TREQ})], which
gives, once we set all the scalene parameters to zero, the
corresponding double-parenthesis symbol [of the kind of
Eq.(\ref{FINAL})].

What about the equation with 2 and 4 different overlaps?  When we set
all the scalene parameters to zero, we have that the expressions
$E^{(i,j)}$ are the sum of $k\!-\!2$ of these double-parenthesis
symbols, those derived from the overlap terns $(q_i,q_j,q_h)$ with
$q_h \ne q_i$ and $q_h \ne q_j$ (that is those where the parameter
$a'_{ij}$ appears).  On the other hand the equations with 4 different
overlaps are identically satisfied and they are useless.

Then we conclude that, in the more general solution, Eq.(\ref{FINAL})
must hold for every overlaps tern $q_i<q_j<q_l$.  What does it imply
in terms of the ultrametric properties of \PPP?

For any pair of overlaps $q_i > q_j$ we have two different isosceles
triangles: a ``right'' one (i.e.\ allowed by UM) with probability
$p_{ijj}$ and a ``wrong'' one (i.e.\ forbidden by UM) with probability
$p_{iij} \propto a'_{ij}$.  In the solution we have found, almost all
the wrong isosceles triangles are forbidden.  More precisely, for any
pair of overlaps $q_i > q_j$, such that there is an overlap $q_h$ in
between ($q_i > q_h > q_j$), we have that $a'_{ij}=0$ and the wrong
isosceles triangle is not allowed.  That can be easily proved noting
that, for any $q_i > q_h > q_j$, the equation
$((a'_{ij};a'_{ih},a'_{hj})) = 0$ forces $a'_{ij}=0$.

Small UM violations can appear only when one considers nearest
neighbors overlaps pairs.  In this case both the right and the wrong
isosceles triangles are allowed.  However, for any fixed $k$, the
maximum number of wrong isosceles triangles allowed is
$[\frac{k}{2}]$, while the total number of isosceles triangles is
proportional to $k^2$.  So in the limit $k \to \infty$ the probability
of having wrong isosceles triangles tends to zero.

Moreover, if in the continuum limit the $P(q)$ is dense on a single
compact domain, the distance between any pair of nearest neighbors
overlaps tends to zero for $k \to \infty$ and then strict UM holds for
any finite overlap difference $|q_i-q_j|$.

Lastly it should be noted that we have not exploited all the available
information and maybe even these small UM violations could be ruled
out with some more work.  In fact we believe that the solution with
both $p_{iij}$ and $p_{ijj}$ different from zero, actually corresponds
to the case $q_i = q_j$ and it is not really an UM violation.

Maybe it should be also possible to find a direct proof of our result
directly in the continuum limit, without considering the intermediate
case in which the number of steps are finite, using maybe the
techniques introduced by Ruelle~\cite{RUELLE}, however we have not
succeeded in this task.

\section{Conclusions}

We have seen that in systems where the function $P(q)$ is non trivial
and the overlap is a fluctuating quantity, stochastic stability and
separability imply ultrametricity.  The reader should notice that the
proofs presented here are likely to be too involved and it is quite
possible that there is a direct proof of the fact that replica
equivalence implies ultrametricity.  At this end we recall that in a
dynamical approach it was show that one can identify a dynamical
equivalent of separability: i.e.\ we can assume that in the aging
regime all the possible overlaps among two configurations at two quite
different times ($t_{1}$ and $t_{2}$) are functions of the usual
overlap among the two configurations at the same times ($t_{1}$ and
$t_{2}$).  It was possible to prove that this dynamical overlap
equivalence implies a dynamical form of ultrametricity.

This result implies that, if we do not give up stochastic stability
(which is a general property of generic equilibrium systems)
violations of ultrametricity may be found only in systems for which
the separability conditions does not hold and the mutual relations
among equilibrium configurations is described by two or more overlaps.
The probability distribution of such a system (if it exists in the
framework of equilibrium statistical mechanics) would be much more
complex of that of the usual ultrametric Ansatz.  We can thus conclude
that the ultrametric solution is the simplest one.

Our arguments imply that it would be particularly interesting to check
the numerical validity of overlap equivalence.  This task can be done
at high precision using present numerical technology. This could be
done just doing numerical simulations in the ensemble with fixed
overlap and looking to the fluctuations of the other overlaps.

\section*{Acknowledgments}

We thank S. Franz and M. A. Virasoro for many interesting discussion
on the subject of the paper.

\section*{Appendix I}

In this appendix we present some arguments in order to show that {\em
separability} implies {\sl overlap equivalence}.

To this end let us consider a specific spin glass model
\begin{equation}
H_J = \sum_{i,k} J_{ik} \sigma_i \sigma_k \quad ,
\end{equation}
where the variables $J$ are Gaussian uncorrelated random variables
with zero average and variance
\begin{equation}
\overline{J^2_{ik}} = K_{ik} \quad .
\end{equation}
In short range models $K_{ik}$ is a fast decreasing function of the
distance among the two points $i$ and $k$ while in the SK model
$K_{ik}=N^{-1}$, $N$ being the total number of spins.

In this model we can define not only the usual overlap but also a
modified overlap among two configurations $\sigma$ and $\tau$ which we
denote by $r$:
\begin{equation}
r = \frac1N \sum_{i,k} J_{ik} \sigma_i \tau_k \quad .
\end{equation}
In the same way we can define an overlap among two replicas, which we
denote by $r_{ab}$
\begin{equation}
r_{ab} = \frac1N \sum_{i,k} J_{ik} \lan \sigma_i^a \sigma_k^b \ran
\quad .
\end{equation}

We notice that by simple integration by part on the Gaussian variables
$J$ one can prove that
\begin{equation}
\overline{\lan r \ran} = \frac{1}{n(n-1)} \sum_{a,b=1,n} t_{ab} \quad ,
\end{equation}
with
\begin{equation}
t_{ab} = {\lan r_{ab} \ran} \sum_{i,k} K_{ik} \sum_{c}
\frac{1}{n(n-1)} \lan \sigma_i^a \sigma_k^b \sigma_i^c \sigma_k^c \ran
\quad ,
\end{equation}
where by the overbar we denote the average over the random quenched
variables $J$.

A similar computation tells us that the fluctuations of the quantity
$r$ at fixed $q$ are (neglecting terms which go to zero with the
volume) the same as the fluctuations of $t_{ab}$ at fixed $q_{ab}$. On
the other end it is evident that
\begin{equation}
\sum_c \sum_{i,k} \frac{1}{N^2} \lan \sigma_i^a \sigma_k^b \sigma_i^c
\sigma_k^c \ran = \sum_c Q_{ac} Q_{bc} \quad .
\end{equation}
Separability states that the last sum takes a fixed value which does
not fluctuates if we stay in the ensemble of fixed $Q_{ab}$.

At this level the relation among separability and overlap equivalence
is clear: the first is equivalent to the statement that the quantity
$N^{-1}\sum_{i}\sum_{c}\lan \sigma_{i}^{a}\sigma_{k}^{b}
\sigma_{i}^{c}\sigma_{k}^{c}\ran$ does not fluctuate (in the fixed
$q_{ab}$ ensemble) when $i-k$ is large, while for overlap equivalence
we need the same quantities does not fluctuate when the distance among
$i$ and $k$ is fixed.

The two properties seems to be slightly different if the interaction
is short range. On the contrary if the interaction is long range the
two formulations are the same. In order to be more precise we can
consider a model in which a small long range interaction has been
added and the same argument as before can be used to prove that
overlap equivalence implies separability. Moreover the presence of a
long range term should not affect too much the properties according to
the principle of stochastic stability which tells us that the system
should be stable with respect to a small random perturbation.

The argument we have presented here tell us that replica equivalence
implies separability.

\section*{Appendix II}

In this appendix we show some details of the computation we have done
in the cases $k=4$ and $k=5$ .  In particular we show exactly how to
derive the solution for the $k=4$ case, while we simply sketch it in
the $k=5$ case.

\subsection*{The $k=4$ case}

In the case $k=4$ we have 4 scalene parameters ($p_{432}, p_{431},
p_{421}, p_{321}$) and 6 isosceles parameters ($p_{443}=a_{43},
p_{442}=a_{42}, p_{441}=a_{41}, p_{332}=a_{32}, p_{331}=a_{31},
p_{221}=a_{21}$).  The remaining probabilities are functions of these
10 parameters:
\begin{equation}
\begin{array}{l}
p_{433} = p_4 p_3 /2 -a_{43} -p_{432} -p_{431} \\
p_{422} = p_4 p_2 /2 -a_{42} -p_{432} -p_{421} \\
p_{411} = p_4 p_1 /2 -a_{41} -p_{431} -p_{421} \\
p_{322} = p_3 p_2 /2 -a_{32} -p_{432} -p_{321} \\
p_{311} = p_3 p_1 /2 -a_{31} -p_{431} -p_{321} \\
p_{211} = p_2 p_1 /2 -a_{21} -p_{421} -p_{321} \\
p_{444} = p_4(1+p_4)/2 -a_{43} -a_{42} -a_{41} \\
p_{333} = p_3(1+p_3-p_4)/2 -a_{32} -a_{31} +a_{43} +p_{432} +p_{431} \\
p_{222} = p_2(1+p_2-p_3-p_4)/2 -a_{21} +a_{42} +a_{32} +2 p_{432}
+p_{421} +p_{321} \\
p_{111} = p_1^2 +a_{41} +a_{31} +a_{21} +2 (p_{431}+p_{421}+p_{321})
\end{array}
\label{def}
\end{equation}

Let's consider the equation of the kind of Eq.(\ref{DUEQ}) with the
two greatest overlaps ($q_4$ and $q_3$ in the $k=4$ case),
\begin{equation}
E^{(4,3)} = \frac12 p_4 p_3 +\frac{p_{431}^2}{p_1}
+\frac{p_{432}^2}{p_2} +\frac{p_{433}^2}{p_3} +\frac{a_{43}^2}{p_4}
-\left(\frac{a_{41} a_{31}}{p_1} +\frac{a_{42} a_{32}}{p_2}
+\frac{a_{43} p_{333}}{p_3} +\frac{p_{444} p_{433}}{p_4}\right) = 0
\quad .
\label{ciao}
\end{equation}
Using some of Eqs.(\ref{def}) and multiplying the previous equation by
$c = 2 p_1 p_2 p_3 p_4$ we end with the following equation
\begin{eqnarray}
c E^{(4,3)} = c E_0^{(4,3)} + \qquad \qquad \qquad \qquad \qquad
\qquad \qquad \qquad \qquad \qquad \qquad \qquad \qquad \qquad
\nonumber \\
+[2 p_1 p_3 p_4 a_{43} + p_1 p_2 p_3 (p_4 p_3 -2 a_{43} +p_4 p_2 -2
a_{42} +p_4 p_1 -2 a_{41})] (p_{432}+p_{431}) + \nonumber \\
+[4 p_1 p_3 p_4] p_{432} p_{431} +[2 p_1 p_4 (p_2+p_3)] p_{432}^2 +[2
p_2 p_4 (p_1+p_3)] p_{431}^2 = 0 \quad . \quad
\label{eq43}
\end{eqnarray}
In $E_0^{(4,3)}$ there are all the terms that survive from the
expression $E^{(4,3)}$ when we set all the scalene parameters to zero.
In general all the expressions $E_0^{(i,j)}$ are non-negative defined
(see below).

The coefficients of $p_{432}$ and $p_{431}$ in Eq.(\ref{eq43}) are
positive defined, thanks to the inequalities $p_4 p_3 \ge 2 a_{43},
p_4 p_2 \ge 2 a_{42}, p_4 p_1 \ge 2 a_{41}$, that are direct
consequence of Eqs.(\ref{def}) and of probabilities positiveness.
Then we have that Eq.(\ref{eq43}) is equivalent to $p_{432}=p_{431}=0$
and $E_0^{(4,3)}=0$.

To force the two remaining scalene parameters to zero, it is enough to
consider the equation analogous to Eq.(\ref{ciao}) with $q_1$ and
$q_2$ instead of $q_3$ and $q_4$.  Once we set $p_{432}=p_{431}=0$ we
obtain
\begin{eqnarray}
c E^{(2,1)} = c E_0^{(2,1)} +[2 p_1 p_3 p_4 (a_{42}+a_{32}) +p_1 p_2
p_3 (p_4 p_2 -2 a_{42} +p_4 p_1 -2 a_{41})] p_{421} + \nonumber \\
+[2 p_1 p_3 p_4 (a_{42}+a_{32}) +p_1 p_2 p_4 (p_3 p_2 -2 a_{32} +p_3
p_1 -2 a_{31})] p_{321} + \nonumber \\
+[2 p_3 p_4(p_1+p_2)] (p_{421} + p_{321})^2 = 0 \quad .
\label{eq21}
\end{eqnarray}
Again we note that all the coefficient are positive defined thanks to
Eqs.(\ref{def}) and to the positiveness of the probabilities.
Eq.(\ref{eq21}) implies $p_{421}=p_{321}=0$ and $E_0^{(2,1)}=0$.

Then we conclude that the more general solution to Eq.(\ref{MAIN}) in
the case of $k=4$ different overlaps forbids any scalene triangle.

In order to obtain this result we have made only one assumption, about
the non-negativity of $E_0^{(4,3)}$ and of $E_0^{(2,1)}$, which we now
show to be correct.  Using the rescaled variables, $a'_{ij} = \frac{2
a_{ij}}{p_i p_j}$, those expressions read
\begin{eqnarray}
E_0^{(4,3)} = \frac{p_2 p_3 p_4}{4} ((a'_{42};a'_{43},a'_{32})) +
\frac{p_1 p_3 p_4}{4} ((a'_{41};a'_{43},a'_{31})) \quad , \\
E_0^{(2,1)} = \frac{p_1 p_2 p_4}{4} ((a'_{41};a'_{42},a'_{21})) +
\frac{p_1 p_2 p_3}{4} ((a'_{31};a'_{32},a'_{21})) \quad . \,
\end{eqnarray}
Note that $E_0^{(4,3)}$ (resp. $E_0^{(2,1)}$) is the sum of the two
double-parenthesis symbols containing $a'_{43}$ (resp. $a'_{21}$).

Once we set to zero all the scalene parameters
$p_{432}=p_{431}=p_{421}=p_{321}=0$, we found easier to work with
equations of the kind of Eq.(\ref{TREQ}).  For example, considering
the 3 overlaps $q_1<q_2<q_3$, we have that
\begin{equation}
0 = P^{12,23,34,41}(q_1,q_1,q_2,q_3) -
P^{12,23,34,41}(q_1,q_2,q_3,q_1) = \frac{p_1 p_2 p_3}{4}
((a'_{31};a'_{32},a'_{21})) \quad .
\end{equation}
In general for every 3 given overlaps we end with an equation like
Eq.(\ref{FINAL}).

\subsection*{The $k=5$ case}

The way to force the scalene parameters to zero should be now clear:
it exploit the coefficient positiveness in the equations with 2
different overlaps.  Maybe it is still not so clear if there is a
systematic way to set all those parameters to zero, without getting
lost in the many $E^{(i,j)}$ expressions.

We found such a systematic way and we will illustrate it in the case
with $k=5$ different overlaps.  Let's always consider first the
equation with the two greatest overlaps ($q_5$ and $q_4$ in this
particular case).  It implies
\begin{equation}
E^{(5,4)} = 0 \quad \Longrightarrow \quad p_{543}=p_{542}=p_{541}=0
\quad .
\end{equation}
Note that all the scalene probabilities forced to be zero contain both
$q_5$ and $q_4$.

Then let's substitute the just found solution
($p_{543}=p_{542}=p_{541}=0$) into all the other equations and let's
go forward in the same way
\begin{eqnarray}
E^{(5,3)} = 0 \quad &\Longrightarrow& \quad p_{532}=p_{531}=0 \quad , \\
E^{(5,2)} = 0 \quad &\Longrightarrow& \quad p_{521}=0 \quad .
\end{eqnarray}
At this point we end with the same scalene parameters we work with in
the $k=4$ case and then let's follow the same steps as in the previous
section
\begin{eqnarray}
E^{(4,3)} = 0 \quad &\Longrightarrow& \quad p_{432}=p_{431}=0 \quad , \\
E^{(2,1)} = 0 \quad &\Longrightarrow& \quad p_{421}=p_{321}=0 \quad .
\end{eqnarray}

Once all the scalene probabilities have been forced to zero, the
demonstration is straightforward and follows the same way outlined in
the previous sections for the $k=3,4$ cases.

\end{document}